\documentclass[aps,prb,amsfonts,amssymb,twocolumn,amsmath,floatfix]{revtex4-2}

\usepackage[cp1251]{inputenc}
\usepackage[TS1,T2A]{fontenc}
\usepackage[english]{babel}

\usepackage{subfig}
\usepackage{subfloat}
\usepackage{textcomp}
\usepackage{graphicx}
\usepackage{hyperref}
\usepackage{xcolor}
\definecolor{color}{HTML}{0000FF}
\hypersetup{linkcolor=color,urlcolor=color,citecolor=color,filecolor=color,colorlinks=true}

\usepackage{amsmath}
\usepackage[mathscr]{euscript}
\usepackage{amsbsy}
\usepackage{dcolumn}
\usepackage{bm}
\usepackage{bbm}
\usepackage{overpic}
\usepackage{titlesec}
\usepackage{dsfont}
\usepackage{amsthm}
\usepackage{delarray}
\usepackage{blkarray}

\def\Xint#1{\mathchoice
   {\XXint\displaystyle\textstyle{#1}}%
   {\XXint\textstyle\scriptstyle{#1}}%
   {\XXint\scriptstyle\scriptscriptstyle{#1}}%
   {\XXint\scriptscriptstyle\scriptscriptstyle{#1}}%
   \!\int}
\def\XXint#1#2#3{{\setbox0=\hbox{$#1{#2#3}{\int}$}
     \vcenter{\hbox{$#2#3$}}\kern-.5\wd0}}

\def\dashint{\Xint-}

\newcommand{\be}{\begin{equation}}
\newcommand{\ee}{\end{equation}}
\def\bml#1\eml{\begin{multline}#1\end{multline}}
\def\bal#1\eal{\begin{align}#1\end{align}}
\def\bald#1\eald{\begin{aligned}#1\end{aligned}}

\newcommand{\p}{\partial}

\DeclareMathOperator{\Img}{\mathrm{Im}}

\begin{document}

\title{Damped oscillators within the general theory of Casimir and van der Waals forces}

\author{Yu.\,S.~Barash}

\affiliation{Institute of Solid State Physics of the Russian Academy of Sciences,
Chernogolovka, Moscow District, 2 Academician Ossipyan Street, 142432 Russia}


\begin{abstract}
It is demonstrated that the general theory of Casimir and van der Waals forces describes the interaction-induced 
equilibrium thermodynamic potentials of the damped harmonic oscillator bilinearly coupled to the environment.
An extended model for a damped oscillator is suggested along the lines of the general theory of Casimir and van der 
Waals forces, and the corresponding thermodynamic quantities obtained. While the original model involves a heat bath 
consisting of a large number of free oscillators having infinitesimal damping functions, the extended model allows 
any generally admissible frequency and temperature dependent dissipative susceptibilities of the heat bath constituents, 
influenced by the additional dissipative environmental channels that are not directly linked to the system oscillator. 
Consequently, the results obtained are applicable to the frequency and temperature dependent damping function of
the system oscillator.
\end{abstract}

\maketitle

\section{Introduction}
\label{sec: Intro}

Based on the fluctuation-dissipation relations applied to electromagnetic fields in condensed matter, Lifshitz 
successfully solved the problem of Casimir and van der Waals interaction between thick plates separated by an empty gap 
\cite{Lifshitz1955}. In his paper Lifshitz demonstrated how to obtain the unified description of the Casimir and van der
Waals forces between macroscopic bodies with arbitrary frequency dependent permittivities.

With the advent of the quantum-field methods of the many-body theory, the Lifshitz approach was generalized by 
Dzyaloshinskii and Pitaevskii \cite{DzyaloshinskiiPitaevskii1959}, who derived widely applicable general formulas that
describe contribution of the long wavelength fluctuational electromagnetic field, interacting with the condensed matter,
to thermodynamic quantities of inhomogeneous dissipative systems in equilibrium. This made it possible to extend the 
Lifshitz solution for the problem of two solids separated by an empty gap to the case of the interaction across a liquid
film, in particular \cite{DzyaloshinskiiLifshitzPitaevskii1959}. From then on, fundamental ideas and results
of the general theory of Casimir and van der Waals forces \cite{Lifshitz1955,DzyaloshinskiiPitaevskii1959,%
DzyaloshinskiiLifshitzPitaevskii1959,DzyaloshinskiiLifshitzPitaevskii1961} have become the basis for the  modern 
description of thermodynamic quantities influenced by fluctuational electromagnetic interactions
\cite{Lamoreaux2004,Parsegian2006,KlimchitskayaMohideenMostepanenko2009,Bordagetal2009,%
FrenchParsegianPodgorniketal2010,Dalvitetal2011,Woodsetal2016}.

Much later, irrespective of the research referred to above, thermodynamics of the oscillator bilinearly coupled to the
heat bath attracted considerable attention \cite{FordLewisOConnell1985,FordLewisOConnell1988,FordLewisOConnell1988_2,%
FordOConnell2005,HaenggiIngold2006,FordOConnell2007,HoerhammerBuettner2008,HaenggiIngoldTalkner2008,%
IngoldHaenggiTalkner2009,Philbin2012,SprengIngoldWeiss2013,AdamietzIngoldWeiss2014,PhilbinAnders2016,%
KolarRyabovFilip2019,TalknerHaenggi2020}. Quantum and classical dynamics of a damped oscillator coupled with the finite 
strength to the environment and governed by fluctuations have been studied for many years 
\cite{Magalinskii1959,FordKacMazur1965,Ullersma1966_1_4,Zwanzig1973,CaldeiraLeggett1981,Schmid1982,CaldeiraLeggett1983,%
GrabertWeissTalkner1984,RiseboroughHaenggiWeiss1985,FordKac1987,GrabertSchrammIngold1988,FordLewisOConnell1988_3,%
HaenggiIngold2005,Weiss2012}. The damped oscillator is known to be a representative example of dissipative quantum 
systems \cite{Weiss2012,Caldeira2014}, which is of interest within the context of a number of 
problems, in particular those concerned with the quantum Brownian motion \cite{GrabertSchrammIngold1988,%
HaenggiIngold2005}, with the effects of dissipation on quantum tunneling processes \cite{CaldeiraLeggett1983,%
Leggettetal1987} and with some general aspects of statistical mechanics and thermodynamics at strong coupling with the 
heat bath \cite{TalknerHaenggi2020}.

This paper focuses on the comparison of the basic aspects of the two theories and their results, identifies features 
they have in common and the points of substantial overlaps, and extends the model for a damped oscillator to a more 
general structure of the environment along the lines of the general theory of Casimir and van der Waals forces.
 
At first glance, the two fields of research differ from each other substantially. The Zwanzig-Caldeira-Leggett model of 
a damped oscillator \cite{Zwanzig1973,CaldeiraLeggett1983} exemplifies a small quantum system, which demonstrates a
dissipative behavior that occurs due to its interaction with the environment. The model manifests certain properties, 
which are of importance to the quantum dynamics, and has the advantage of comprising the exact solution for simple 
cases. It can also be considered to be a part of more complicated quantum mechanical problems. The environment, which is
made up of a considerable number of free oscillators that interact only with the system oscillator, allows the losses to
be modeled by reducing the problem to one or two degrees of freedom.

As opposed to the damped oscillator model, the general theory of Casimir and van der Waals forces considers - within the
quantum microscopic approach - the long wavelength components of the electromagnetic field that interact with realistic
condensed matter systems, so that there is neither a need for the theory to classify the two subsystems into a small 
central system and a large thermal reservoir, nor any need for a significant simplification. Frequency dependent 
dissipative permittivities of the condensed matter enter the theory in a general unspecified form, enabled by the 
fluctuation-dissipation relations, and are assumed to be provided by independent theoretical studies or experiments. The
general theory applies to inhomogeneous condensed systems of arbitrary spatial profile, that manifest nonlocal 
correlations of the microscopic quantum electromagnetic field in media with spatially dependent matrix permittivities. 
Free from the model-imposed assumptions, the theory offers a significantly more powerful approach within the range of 
its applicability.

On the other hand, the theories have a number of important features in common. Since both Casimir and van der Waals 
interactions and the environment-affected oscillator are governed by equilibrium fluctuations, it is known that the 
theories share some similarities \cite{IngoldLambrechtReynaud2009}, although there has been no direct comparison of 
their basic results for thermodynamic potentials. Furthermore, the long wavelength electromagnetic fluctuational field 
in the theory of Casimir and van der Waals forces manifests a dissipative behavior with complex eigenfrequencies due to 
its interaction with the condensed matter, and in this context resembles the system oscillator that becomes dissipative
due to its interaction with the environment. 

One should also note the common bilinear operator structure of the basic interaction terms in the 
Hamiltonians, with the quantum microscopic current density and electromagnetic field in one of the theories, and 
positions of both the system and environmental oscillators in the other. In both cases this allows to average the 
interaction terms using the fluctuation-dissipation relations. One more significant feature the two theories have in 
common is the fact that the interaction with the system damped oscillator does not modify, within the 
model, the susceptibilities of individual environmental oscillators. This is similar to the general theory of Casimir
and van der Waals forces, which assumes the contribution to the condensed matter permittivities from the interaction 
with the long wavelength fluctuational electromagnetic field to be negligibly small. 

A substantial overlap between the two theories under discussion becomes apparent, when the damped oscillator's 
interaction with the environment is of the electromagnetic origin. Having taken this point as a given, and having 
conducted its analysis, to an extent, in the reversed order, the free energy induced by the Casimir and van der 
Waals interaction in dissipative systems was identified in Ref. \cite{BarashGinzburg1972} on the basis of the study of 
long wavelength electromagnetic fluctuations in the elementary RCL circuit, that exemplifies the damped oscillator. 
Although, while the general theory is applicable to the conventional RCL circuit, the circuit does not fully match the
conditions of the model, in which the environmental oscillators have infinitesimal damping functions. By contrast, 
the conventional resistance of the RCL circuit is dominated in the metal by some real processes that should be 
considered as additional internal dissipative channels of the environment not included in the model. This, of course, 
relates to most of the systems usually studied with the general theory of Casimir and van der Waals forces
\cite{footnote}.
 
A model of the electromagnetic origin that simultaneously satisfies the conditions of both the model of a damped 
oscillator and the theory of Casimir and van der Waals forces, is that of the charged oscillator in the blackbody 
radiation field, where the oscillator's damping function is produced by the radiation reaction processes 
\cite{FordLewisOConnell1985,FordLewisOConnell1988}. The relation of this particular problem to the theory of Casimir and
van der Waals interaction, omitted in \cite{FordLewisOConnell1985,FordLewisOConnell1988}, was discussed in 
\cite{Obcemea1987} but only for the simplest limit of real eigenfrequencies. In the latter case, the result is reduced 
to the free energy of free oscillators with eigefrequencies of the interacting system, as is the case in describing 
Casimir and van der Waals interaction in systems with real eigenfrequencies \cite{FLondon1937,Casimir1948,%
vanKampenNijboerSchram1968,NinhamParsegianWeiss1970,BarashGinzburg1975}.

The interaction-induced thermodynamic potentials, obtained both with the model for a damped oscillator and with the 
model-free general theory of Casimir and van der Waals forces, will be found to coincide, when they are expressed in 
terms of similar quantities. It is the consequence of a significant overlap of the two approaches. In other words, 
thermodynamics of a damped oscillator bilinearly coupled to the heat bath is directly described by the results obtained
in the general theory of Casimir and van der Waals forces long ago \cite{DzyaloshinskiiLifshitzPitaevskii1961,%
AbrikosovGorkovDzyaloshinskii1965,Lifshitz1995,BarashGinzburg1975,Barash1988,BarashGinzburg1989}.

An extension of the original model for a damped oscillator conducted below includes additional internal dissipative 
channels of the environment. As a result, instead of the susceptibilities of the free environmental oscillators of the 
original model, the extended model allows for generally admissible frequency and temperature dependent dissipative
susceptibilities. Although it is assumed that the system oscillator is not directly linked with the additional channels,
it acquires a frequency and temperature dependent damping function via the bilinear interaction with dissipative 
environmental oscillators or similar constituents. The extended model allows thermodynamic description of the damped 
oscillator in equilibrium, and is generally not intended to study the total system's dynamics. The corresponding free 
energy, obtained with the temperature dependent damping function of the system oscillator, is shown to coincide with the
free energy obtained with the original model, while the results for the internal energy have been modified.

The paper is organized as follows: Section \ref{sec: model} introduces basic results for the damped oscillator used in 
the sections that follow. Section \ref{sec: freeen} presents an alternative derivation of the interaction-induced free 
energy and the internal energy of the damped oscillator, which is the basis for further consideration. Sections 
\ref{sec: gen1}-\ref{sec: gen3} reveal connections of the results obtained for the damped oscillator to the results of 
the general theory of Casimir and van der Waals interaction. The extended model of a damped oscillator is developed in 
Section \ref{sec: freeenmod}, where the corresponding thermodynamic potentials are obtained as well. Section 
\ref{sec: conclusions} concludes the paper.

\section{Basic equations for a damped oscillator coupled to a heat bath}
\label{sec: model}

Let a small system coupled to the heat bath be the system oscillator in contact with environmental oscillators, 
described by position operators $\hat Q$ and $\hat q_\alpha$($\alpha=1,\,2,\,\ldots,N$), respectively.
If the external perturbation in the Hamiltonian is of the form
\be
\hat V_{\text{ext}}=-\hat{Q}f_{\text{ext},Q}(t)-\sum_{\alpha=1}^N\hat{q}_\alpha f_{\text{ext},\alpha}(t),
\label{f1}
\ee
and the linear response to the external forces $f_{\text{ext},Q}$, $f_{\text{ext},\alpha}$ is described by the following
relations for the Fourier components of the averaged quantities
\bal
&Q(\omega)=\chi_{QQ}(\omega)f_{\text{ext},Q}(\omega)+
\sum_{\alpha=1}^N\chi_{Q\alpha}(\omega)f_{\text{ext},\alpha}(\omega),
\label{f2}\\
&q_\alpha(\omega)=\chi_{\alpha Q}(\omega)f_{\text{ext},Q}(\omega)+
\sum_{\delta=1}^N\chi_{\alpha\delta}(\omega)f_{\text{ext},\delta}(\omega),
\label{f3}
\eal
then the fluctuation-dissipation theorem is known to read (see, for example, \cite{LandauLifshitz5_1980})
\bal
&\bigl(Q^2\bigr)_{\omega}=\hbar\coth\dfrac{\hbar\omega}{2T}\Img\Bigl[\chi_{QQ}(\omega)\Bigr],
\label{f4} 
\\
&\bigl(q_\alpha Q\bigr)_{\omega}=\dfrac{i\hbar}2\coth\dfrac{\hbar\omega}{2T}
\Bigl[\chi_{Q\alpha}^*(\omega)-\chi_{\alpha Q}(\omega)\Bigr],
\label{f5} \\
&\bigl(q_\alpha^2\bigr)_{\omega}=\hbar\coth\dfrac{\hbar\omega}{2T}
\Img\Bigl[\chi_{\alpha\alpha}(\omega)\Bigr],
\label{f6}\\
&\bigl(q_\alpha q_\delta\bigr)_{\omega}=\dfrac{i\hbar}2\coth\dfrac{\hbar\omega}{2T}
\Bigl[\chi_{\delta\alpha}^*(\omega)-\chi_{\alpha\delta}(\omega)\Bigr].
\label{f7}
\eal

Here $\bigl(AB\bigr)_{\omega}$ denotes the spectral function of the symmetrized correlation function in the
equilibrium state
\be
\dfrac12\langle\hat{A}(t)\hat{B}(t')+\hat{B}(t')\hat{A}(t)\rangle=
\int\limits_{-\infty}^{\infty}(AB)_\omega e^{-i\omega(t-t')}\dfrac{d\omega}{2\pi}.
\label{f8}
\ee

If the quantum states are invariant with respect to the time inversion, the relations $\chi_{Q\alpha}(\omega)
=\chi_{\alpha Q}(\omega)$, $\chi_{\alpha\delta}(\omega)=\chi_{\delta\alpha}(\omega)$ will hold 
and, for example, 
Eq. \eqref{f5} reduces to the relation $\bigl(q_\alpha Q\bigr)_{\omega}=\hbar\coth\frac{\hbar\omega}{2T}
\Img\bigl[\chi_{Q\alpha}(\omega)\bigr]$.

The linearized dynamic equation for a statistically averaged position of the system oscillator of mass $M$ and frequency 
$\Omega$ is, in the presence of the environment-induced dissipation and an external force,
\be
M\ddot{Q}(t) + M \!\! \int\limits_{-\infty}^{t}\!\! dt'\tilde{\gamma}(t-t')\dot{Q}(t') + M\Omega^2Q(t)=f_{\text{ext}}(t),
\label{f9}
\ee
where $\tilde{\gamma}(t)$ is the dissipative memory function.

The corresponding frequency dependent susceptibility, therefore, takes the form $\chi_{QQ}(\omega)\equiv
\chi_{Q}(\omega)$,
\be
\chi_{Q}(\omega)=\dfrac{1}{M\left[\left(\Omega^2-\omega^2\right)-i\omega \gamma(\omega)\right]},
\label{f10}
\ee
where the damping function $\gamma(\omega)$ is the Fourier transform of $\tilde{\gamma}(t)$.

The susceptibility of the $\alpha$th free oscillator with mass $m_\alpha$ and frequency $\omega_\alpha$ is
\be
\chi_{\alpha}(\omega)=\dfrac{1}{m_\alpha}\dfrac{1}{\omega_\alpha^2-\omega^2-i\omega\varepsilon}, 
\enspace \varepsilon\to+0.
\label{f11}
\ee

For specifying crossed components $\chi_{Q\alpha}(\omega)$, $\chi_{\alpha\gamma}(\omega)$ of the matrix response 
function it is now convenient to introduce the full quantum Hamiltonian of the Zwanzig-Caldeira-Leggett
model for a damped oscillator \cite{Zwanzig1973,CaldeiraLeggett1983} with the classical external forces added:
\bml
\hat{H}=\dfrac{\hat{P}^2}{2M}+\dfrac12 M\Omega^2 \hat{Q}^2-\hat{Q}f_{\text{ext},Q} +
\sum_{\alpha=1}^N\left[\dfrac{\hat{p}_\alpha^2}{2m_\alpha} \right.\\ \left.+
\dfrac12m_\alpha\omega_\alpha^2
\left(\hat{q}_\alpha-\dfrac{C_\alpha}{m_\alpha\omega_\alpha^2}\hat{Q}\right)^2\right]-
\sum_{\alpha=1}^N\hat{q}_\alpha f_{\text{ext},\alpha}.
\label{f12}
\eml

Here $\hat Q$, $\hat P$ and $\hat{q}_\alpha$, $\hat{p}_\alpha$ are position and momentum operators of the system 
oscillator and the $\alpha$th oscillator, respectively. The system oscillator interacts with the $\alpha$th 
environmental oscillator with the coupling strength $C_\alpha$. The interaction terms in 
\eqref{f12} enter the Hamiltonian in the form that excludes the interaction-induced renormalization of the frequency 
$\Omega$.

Comparatively simple model \eqref{f12} is known to allow a detailed description of the damped oscillator influenced by 
the environment \cite{Zwanzig1973,CaldeiraLeggett1983,GrabertSchrammIngold1988,HaenggiIngold2005,Weiss2012,%
TalknerHaenggi2020}. In particular, the quantum equations of motion, after excluding $\hat{q}_\alpha$ from the equation 
for $\hat{Q}$, performing statistical averaging and the Fourier transform, reduce to
\bal
M\left[\left(\Omega^2-\omega^2\right)-i\omega \gamma(\omega)\right]Q(\omega)=& f_{\text{ext},Q}(\omega) \nonumber\\ 
+ \sum_{\alpha=1}^N C_\alpha&\chi_\alpha(\omega)f_{\text{ext},\alpha}(\omega), \label{f13}\\
m_\alpha(\omega_\alpha^2-\omega^2)q_\alpha(\omega)-C_\alpha Q(\omega)=&f_{\text{ext},\alpha}(\omega).
\label{f14}
\eal

One gets from \eqref{f13}, \eqref{f14}
\bal
&\chi_{QQ}(\omega)=\chi_{Q}(\omega), 
\label{f15}\\
&\chi_{Q\alpha}(\omega)=\chi_{\alpha Q}(\omega)=C_\alpha\chi_Q(\omega)\chi_\alpha(\omega), 
\label{f16}\\
&\chi_{\alpha\alpha}(\omega)=\chi_\alpha(\omega)\Bigl(1+C_\alpha^2\chi_Q(\omega)\chi_\alpha(\omega)\Bigr), 
\label{f17}\\
&\chi_{\alpha\delta}(\omega)=C_\alpha C_\delta\chi_{Q}(\omega)\chi_\alpha(\omega)\chi_\delta(\omega),\enspace 
\alpha\ne\delta,
\label{f18}
\eal
where $\chi_{Q}(\omega)$ and $\chi_\alpha(\omega)$ are defined in \eqref{f10}, \eqref{f11}.

As seen in \eqref{f4}-\eqref{f7}, \eqref{f15}-\eqref{f18} and \eqref{f10}, \eqref{f11}, the correlation functions
considered contain the only quantity $\gamma(\omega)$, which is specified within the model \eqref{f12} as
\be
i\omega M\gamma(\omega)=\sum_{\alpha=1}^N C_\alpha^2 \left[\chi_\alpha(\omega)-\chi_\alpha(0)\right]
\label{f19} 
\ee
and can be linked with the so-called interaction spectral density
\be
J(\omega)=\theta(\omega)\sum_{\alpha=1}^N C_\alpha^2\chi''_\alpha(\omega),
\label{f20} 
\ee
where $\theta(\omega)$ is the unit step function and the double prime for $\chi$ signifies the imaginary part.

The following expression for the damping function $\gamma(\omega)$ via the interaction 
spectral density takes place
\be
\gamma(\omega)=\dfrac{-2i\omega}{\pi M}\!\int_{0}^{\infty}\!\!
\dfrac{J(\xi)d\xi}{\xi(\xi^2-\omega^2-i\omega\varepsilon)}, \enspace \varepsilon\to+0.
\label{f21}
\ee

The quantities defined in \eqref{f19}, \eqref{f20} satisfy \eqref{f21} not only when the particular expression 
\eqref{f11} is used, but also for any admissible frequency dependent susceptibilities $\chi_\alpha(\omega)$. For 
establishing the validity of \eqref{f21} in a general form, one should substitute \eqref{f19} and \eqref{f20} in 
\eqref{f21}, and apply to expressions $\bigl[\chi_\alpha(\omega)-\chi_\alpha(0)\bigr]/\omega^2$ under the summation sign
in the left-hand side the following Kramers-Kronig relation 
\be
\chi'(\omega)=\dfrac{2}{\pi}\dashint\limits_0^\infty\dfrac{\xi\chi''(\xi)d\xi}{\xi^2-\omega^2}.
\label{f22}
\ee

\section{Thermodynamic potentials of the damped oscillator}
\label{sec: freeen}

A comparatively simple way to find the interaction-induced contribution to the free energy of the system oscillator 
coupled to the heat bath is to equate, based on the general relationships of statistical physics 
\cite{LandauLifshitz5_1980,Abrikosov1988,Barash1988,BarashGinzburg1989}, the free-energy derivative over the coupling 
parameter to the corresponding averaged derivative of the Hamiltonian, followed by the integration over the coupling 
parameter. To this end, multiplying all coupling constants $C_\alpha$($\alpha=1,\,2,\,\ldots,N$) in the Hamiltonian 
\eqref{f12} by the same parameter $\lambda$, one gets for the averaged derivative of the Hamiltonian over $\lambda$: 
\bml
\biggl\langle\frac{\p\hat{H}(\lambda)}{\p\lambda}\biggr\rangle_\lambda\!\! =
-\sum_{\alpha=1}^NC_\alpha \!
\int\limits_{-\infty}^{\infty} \!\dfrac{d\omega}{2\pi}\Bigl(\left(q_\alpha Q\right)_{\omega,\lambda}\\ -
\lambda C_\alpha\chi_\alpha(0)\left(Q^2\right)_{\omega,\lambda}\Bigr).
\label{f23}
\eml
Here $<\ldots>_\lambda$ and $(\ldots)_{\omega,\lambda}$ designate the averaging over an equilibrium state of the system with the
Hamiltonian $\hat{H}(\lambda)$ and the corresponding symmetrized spectral density, respectively.

As the susceptibilities \eqref{f11} of the environmental oscillators $\chi_\alpha(\omega)$ do not contain $C_\alpha$, 
they are independent of the coupling parameter $\lambda$, while the dissipative damping function
$\gamma(\omega,\lambda)$, as follows from \eqref{f19}, manifests the quadratic dependence: $\gamma(\omega,\lambda)
=\lambda^2\gamma(\omega)$. The parameter-dependent response function $\chi_Q(\omega,\lambda)$ of the system oscillator
is therefore obtained from \eqref{f10} after substituting $\lambda^2\gamma(\omega)$ for $\gamma(\omega)$.

Applying now the fluctuation-dissipation relationships \eqref{f4}, \eqref{f5} and taking into account \eqref{f15}, 
\eqref{f16}, one expresses the quadratic and bilinear in position operators spectral densities in \eqref{f23} via the 
corresponding response functions and gets the free energy in the form
\bml
\Delta_{\lambda}F=F(\lambda=1)-F(\lambda=0)=\int\limits_0^1 d\lambda \left\langle\dfrac{\p\hat{H}(\lambda)}{\p\lambda}
\right\rangle_\lambda\\ 
=-\Img\int_{-\infty}^{\infty}\dfrac{d\omega}{2\pi}\hbar\coth\dfrac{\hbar\omega}{2T}\int\limits_0^1
\dfrac{i\omega\lambda d\lambda\gamma(\omega)}{\left(\Omega^2-\omega^2\right)-i\omega\lambda^2\gamma(\omega)}\\ =
\Img\!\int_{-\infty}^{\infty}\!\!\dfrac{d\omega}{4\pi}\hbar\coth\dfrac{\hbar\omega}{2T}
\ln\dfrac{\Omega^2-\omega^2-i\omega\gamma(\omega)}{\Omega^2-\omega^2-i\omega\varepsilon}, \enspace \varepsilon\to+0.
\label{f24}
\eml

Based on the analyticity of response functions in the upper half-plane of complex frequency one transforms \eqref{f24} 
to the following form
\be
\Delta_{\lambda}F=F-F_0=T\sideset{}{'}\sum_{n=0}^{\infty}
\ln\dfrac{\Omega^2+\omega_n^2+\omega_n\gamma(i\omega_n)}{\Omega^2+\omega_n^2},
\label{f25}
\ee
where $\omega_n=2\pi n T/\hbar$ is the Matsubara frequency and the prime for the summation sign signifies that the term
with $n=0$ is taken with a half-weight. The condition $\gamma(\omega)\to0$ at $\omega\to\infty$, which is to be 
satisfied in realistic systems, has been taken into account in obtaining \eqref{f25}.

Albeit the term with $n=0$ actually vanishes, it is convenient to retain it in \eqref{f25} for compairing further with 
the analogous expression applied to a more general case. It is a peculiarity of the 
Zwanzig-Caldeira-Leggett and related models that the interaction-induced effects in a system with a single degree of 
freedom vanish in the classical limit. This in particular takes place due to the absence of the interaction-induced 
renormalization of the oscillator's frequency $\Omega$ within the model. 

The interaction-induced part $\Delta_{\lambda}F$ of the free energy of the system oscillator in \eqref{f25} represents 
the difference of the total system's free energies taken with and without the system oscillator's interaction with the 
environment. The free energy $F$ of the damped oscillator, as an open system coupled with the heat bath, is obtained 
from \eqref{f25} by adding to its right-hand side the terms, which are independent of the coupling constants and ensure 
that in the absence of the interaction the quantity $F$ coincides with the free energy of the free harmonic oscillator 
$T\ln\Bigl(2\sinh\frac{\hbar\Omega}{2T}\Bigr)=T\ln\Bigl[\frac{\hbar\Omega}{T}\prod\limits_{n=1}^{\infty}
\Bigl(1+\frac{\Omega^2}{\omega_n^2}\Bigr)\Bigr]$.
Therefore, one gets from \eqref{f25} the following damped oscillator's free energy 
\be
F=T\ln\left[\dfrac{\hbar\Omega}{T}
\prod\limits_{n=1}^{\infty}\left(1+\dfrac{\Omega^2}{\omega_n^2}+\dfrac{\gamma(i\omega_n)}{\omega_n}\right)\right].
\label{f26}
\ee

The free energy of the open system should take the form $F=-T\ln Z(T)$, where $Z$ is the reduced partition function, 
which is the ratio of the partition function of the total system to the one of the unperturbed reservoir.
One obtains, therefore, from \eqref{f26} the well-known reduced partition function of the damped linear quantum 
oscillator \cite{Weiss2012,TalknerHaenggi2020} 
\be
Z(T)=\dfrac{T}{\hbar\Omega}\prod\limits_{n=1}^{\infty}\dfrac{\omega_n^2}{\Omega^2+\omega_n^2+\omega_n\gamma(i\omega_n)}.
\label{f27}
\ee

The expression for the internal energy of the damped oscillator follows directly from \eqref{f26} on account of the 
standard relationship $E=-T^2\left(\frac{\p}{\p T}\frac{F(T)}{T}\right)$:
\be
E=T\sideset{}{'}\sum_{n=0}^{\infty}\dfrac{2\Omega^2+\omega_n\gamma(i\omega_n)-
\omega_n^2\dfrac{d\gamma(i\omega_n)}{d\omega_n}}{\Omega^2+\omega_n^2+\omega_n\gamma(i\omega_n)}.
\label{f28} 
\ee

Formula \eqref{f26} for the free energy does not contain the frequency derivative of the damping function, since
the coupling parameter in \eqref{f23}, \eqref{f24} has been considered to vary at a fixed temperature and, hence,
at fixed Matsubara frequencies. On the other hand, the expressions for those thermodynamic potentials that correspond 
to the processes with varying temperature, should inevitably include the derivative of the frequency dependent damping 
function taken at Matsubara frequencies, as exemplified in \eqref{f28}. 

Internal energy \eqref{f28} coincides with the result previously obtained by using the precalculated reduced partition 
function \cite{HaenggiIngold2006,Weiss2012}. The presence of the frequency derivative of the damping function in 
\eqref{f28} is known to be associated with the temperature dependence of the Hamiltonian of mean force 
\cite{TalknerHaenggi2020}.

\section{Connection to the results by Dzyaloshinskii and Pitaevskii}
\label{sec: gen1}

The aim of the present and the following sections is to compare the basic formulas for thermodynamic potentials, 
obtained in the two theories under consideration, and to demonstrate that they can be presented in the same form when 
written in terms of similar quantities.

This section is based on one of the main model-free results, obtained by Dzyaloshinskii and Pitaevskii in 
Ref.\cite{DzyaloshinskiiPitaevskii1959} within the general theory of Casimir and  van der Waals forces. It represents 
the variation of the interaction-induced free energy of the long wavelength electromagnetic fluctuational field in the 
inhomogeneous condensed matter system.

The linear electromagnetic response of condensed matter that enters the general expressions, is described in the theory
in terms of the polarization operator ${\mathscr P}$, which is taken at Matsubara frequencies and linked with the 
dielectric function as
\be
{\mathscr P}_{ik}(\omega_n,\mathbf{r}_1,\mathbf{r}_2)=\dfrac{\omega_n^2}{\hbar c^2}
\bigl[\epsilon_{ik}(i\omega_n,\mathbf{r}_1,\mathbf{r}_2)-\delta_{ik}\delta(\mathbf{r}_1-\mathbf{r}_2)\bigr].
\label{f29} 
\ee

The corresponding variation of the free energy with a small change of the polarization operator $\delta{\mathscr P}$ has
been found to take the form \cite{DzyaloshinskiiPitaevskii1959}
\be
\delta F=-\dfrac{T}{4\pi}\sideset{}{'}\sum_{n=0}^{\infty}\int d\mathbf{r}_1 d\mathbf{r}_2
{\mathscr D}_{ik}(\omega_n,\mathbf{r}_1,\mathbf{r}_2)\delta{\mathscr  P}_{ki}(\omega_n,\mathbf{r}_2,\mathbf{r}_1).
\label{f30}
\ee

Here ${\mathscr D}_{ik}(\omega_n,\mathbf{r}_1,\mathbf{r}_2)$ is the temperature Green function of the long wavelength 
electromagnetic field in the medium. The corresponding retarded Green function is related to equilibrium correlations of
the microscopic quantum electromagnetic potentials in the medium, with vanishing scalar potential as the gauge 
condition.

Consider now the variation of the interaction-induced free energy \eqref{f25} (or \eqref{f26}) with a small change of 
the damping function:
\be
\delta F=-T\sideset{}{'}\sum_{n=0}^{\infty}\chi_Q(i\omega_n)\bigl[-M\omega_n\delta\gamma(i\omega_n)\bigr].
\label{f31}
\ee
Here the system oscillator response function \eqref{f10} has been used. Eq.\eqref{f31} holds for any possible frequency
dependence of the damping function.

A close similarity of formulas \eqref{f30} and \eqref{f31} can be specified in detail. While the Green function in 
\eqref{f30} is related to the correlations of fluctuating microscopic electromagnetic potentials in the medium, the 
response function $\chi_Q$ describes the correlations of the system oscillator's position $Q$ (see \eqref{f4} and 
\eqref{f15}). Also, the spectral density of electromagnetic Langevin forces in media is known to be related to the 
polarization operator, while the quantity $i\omega M\gamma(\omega)$ is related to the spectral density of the Langevin 
forces in the case of the damped oscillator. 

Furthermore, as seen in \eqref{f13} and \eqref{f10}, the function $\chi_Q(\omega)$ describes the linear response of the
position $Q$ to the external force $f_{\text{ext},Q}(\omega)$, while the relation between the friction force and the 
position $Q$ is $f_{\text{friction}}(\omega)=i\omega M\gamma(\omega)Q(\omega)$. At the same time, the quantity 
$-(1/\hbar c){\mathscr D}$ is known to play the role of a generalized matrix nonlocal response function for the 
electromagnetic potentials induced by external currents, while $-(c\hbar/4\pi){\mathscr  P}$ enters the relation between
the current density and the electromagnetic potential in media. After grouping the quantities in question with indicated
coefficients, one gets the same multipliers in front of the sums in both expressions \eqref{f30} and \eqref{f31}.

Finally, the interaction with the long wavelength electromagnetic field is considered in the general theory as the
interaction with a comparatively weak field in the sense that the contribution from such a field to the permittivity
of the condensed matter is to be negligibly small. Therefore, the polarization operator \eqref{f29} enters the 
expressions of the theory being multiplied by the corresponding coupling constant squared. Similarly, the 
susceptibilities of individual environmental oscillators in the model are assumed to be independent of the relevant 
coupling constants. This results in the quadratic dependence of the damping function $\gamma(\omega)$ on the 
coupling parameter.

\section{Connection to the Casimir and van der Waals component of the total free energy}
\label{sec: gen2}

Going back to the free energy \eqref{f25}, one focuses on two functions $D(\omega)=\Omega^2-\omega^2-
i\omega\gamma(\omega)$ and $D_0(\omega)=\Omega^2-\omega^2-i\omega\varepsilon$ ($\varepsilon\to+0$), taken in \eqref{f25}
at the imaginary Matsubara frequencies $\omega\to i\omega_n$. Roots of the dispersion function $D(\omega)$, i.e., 
solutions of the dispersion equation
\be
D(\omega)=\Omega^2-\omega^2-i\omega\gamma(\omega)=0
\label{f32}
\ee 
are complex eigenmodes of the damped oscillator. At the same time, the equality $D_0(\omega)=0$ defines the eigenmodes 
of a free oscillator, i.e., in the absence of any interaction with the thermal reservoir.

Thus, formula \eqref{f25} takes the form
\be
\Delta_{\lambda}F=F-F_0=T\sideset{}{'}\sum_{n=0}^{\infty}
\ln\dfrac{D(i\omega_n)}{D_0(i\omega_n)}.
\label{f33}
\ee

It should be noted, that Eq. \eqref{f33} represents the basic formula for the free energy in the general theory of 
Casimir and van der Waals forces \cite{BarashGinzburg1975}. The formula can be obtained within the general theory based,
in particular, on the method of integration over the interaction parameter \cite{Barash1988,BarashGinzburg1989}, similar
to that used in Section \ref{sec: freeen} for the problem of the damped oscillator. 

Since Eqs. \eqref{f30} and \eqref{f31} are closely related, and the expression \eqref{f25} coincides with \eqref{f33}, 
one can conclude that the interaction-induced part of the free energy of the damped oscillator is described within the 
framework of the general theory of Casimir and van der Waals forces.

On account of vanishing interaction at infinitely large distances, the parameter $\lambda$ in studying the Casimir and 
van der Waals interaction between the bodies can ultimately be associated with the distance between them. In such a case
the free energies $F$ and $F_0$, as well as the dispersion functions $D(\omega)$ and $D_0(\omega)$ in \eqref{f33}, 
correspond, respectively, to the distance $l$ between the bodies and to the case with the bodies being extremely far 
apart. Therefore, in the limit $l\to\infty$ one has $D(\omega)/D_0(\omega)\to1$. If the function $D(\omega)$ is 
factorized into a number of dispersion functions, then the factors independent of $l$ will cancel each other out in 
\eqref{f33}. The remaining ratio of the two functions is usually considered as a normalized dispersion function that 
provides a distance-dependent spectrum of eigenmodes.

Furthermore,  the eigenmodes in a system of macroscopic bodies of various geometries can depend on the variables that 
run over a continuous or quasi-continuous spectrum of values, such as the components of the wavevector. Separating out 
the continuous variables and labeling them by the letter $\beta$, one obtains from \eqref{f33} the following form for 
the Casimir and van der Waals component of the free energy 
\be
\Delta_{\lambda}F=F-F_0=T\sideset{}{'}\sum_{n=0}^{\infty}\int\rho(\beta)d\beta\ln D(\beta,i\omega_n),
\label{f34}
\ee
where $\rho(\beta)$ is the state density.

Formulas \eqref{f25} or \eqref{f26} for the free energy of the damped oscillator have not been discussed in the
literature, despite their evident connection with the well-known reduced partition function \eqref{f27}. Instead,
the modified form of the result is usually used, which can be obtained from \eqref{f25} by first transforming it to 
the integration along the real frequency axis (cf. \eqref{f24}), using the equality $D(\beta,-\omega)=D^*(\beta,\omega)$ 
and integrating by parts. This leads to the following equivalent to \eqref{f25} (\eqref{f33}) expression:
\bal
&\Delta_{\lambda}F=\int_{0}^{\infty}T\ln\left(2\sinh\dfrac{\hbar\omega}{2T}\right)\Delta_{\lambda}\rho(\omega)d\omega,
\label{f35}\\
&\Delta_{\lambda}\rho(\omega)=\rho-\rho_0=-\dfrac1{\pi}\Img \dfrac{\p}{\p\omega}\ln\dfrac{D(\omega)}{D_0(\omega)}.
\label{f36}
\eal
The partial derivative over frequency instead of the total one implies here possible appearance of additional 
variables that the dispersion function can depend upon. This possibility is discussed in the end of the present section 
and in Section \ref{sec: freeenmod}.

Unlike the theory of Casimir and van der Waals interaction that in most cases considers only the interaction-induced 
part of the free energy, the simple model of the damped oscillator enables us to identify the free energy \eqref{f26} of
the damped oscillator interacting with the environment. Rewriting \eqref{f26} in the form similar to \eqref{f35}, 
\eqref{f36}, one gets with $D(\omega)$ defined in \eqref{f32}
\bal
&F=\int_{0}^{\infty}T\ln\left(2\sinh\dfrac{\hbar\omega}{2T}\right)\rho(\omega)d\omega,
\label{f37}\\
&\rho(\omega)=-\dfrac1{\pi}\Img \dfrac{\p}{\p\omega}\ln \bigl(MD(\omega)\bigr).
\label{f38}
\eal

In view of the equality $MD(\omega)=\chi_Q^{-1}(\omega)$, that follows from \eqref{f10} and \eqref{f32}, 
one obtains from Eqs.\eqref{f37}, \eqref{f38} within the original model of the damped oscillator
\bal
&F=\dfrac1{\pi}\int_{0}^{\infty}T\ln\left(2\sinh\dfrac{\hbar\omega}{2T}\right)\Img \dfrac{d\ln \chi_Q(\omega)}{d\omega}
d\omega,
\label{f39}
\eal
that coincides exactly with the result for the free energy used in the theory of a 
damped oscillator bilinearly coupled with the heat bath \cite{FordLewisOConnell1985,FordLewisOConnell1988,%
FordLewisOConnell1988_2,FordOConnell2005,FordOConnell2007,TalknerHaenggi2020}. 

Similarly to Eqs.\eqref{f35}, \eqref{f36}, the interaction-induced free energy \eqref{f34} of the general theory of 
Casimir and van der Waals forces can also be rewritten as
\bal
&\Delta_\lambda F=\int_{0}^{\infty}T\ln\left(2\sinh\dfrac{\hbar\omega}{2T}\right)\Delta_\lambda\rho(\omega)d\omega,
\label{f40}\\
&\Delta_\lambda\rho(\omega)=-\dfrac1{\pi}\Img \dfrac{\p}{\p\omega}\int\rho(\beta)d\beta\ln D(\beta,\omega).
\label{f41}
\eal

Either of the two formulas \eqref{f26}, \eqref{f39} can, generally speaking, be used for describing the free energy of 
the damped oscillator. Since \eqref{f26} has a simpler form and does not contain the frequency derivative of the 
oscillator's susceptibility, the free energy in the form \eqref{f26} or \eqref{f33} seems to be more preferable as 
compared to \eqref{f39} or \eqref{f37}, \eqref{f38}. This is especially true with respect to the analysis of more 
complicated problems and in particular of typical problems of the theory of Casimir and van der Waals interactions, 
where the free energy in the form \eqref{f34} has a number of significant advantages over \eqref{f40}, \eqref{f41}. 

As known, permittivities take frequency dependent complex values and can manifest a complicated behavior along the real 
frequency axis. By contrast, they take real values and show a comparatively simple monotonic behavior along the upper 
imaginary half-axis. That was Lifshitz's reason \cite{Lifshitz1955} for transforming his basic result, initially 
including the integration over real frequencies, to the form containing the sum over Matsubara frequencies along the 
imaginary upper half-axis \cite{footnote2}.

In addition, the derivatives of permittivities over the frequency that arise after applying \eqref{f41} to inhomogeneous
condensed matter systems, can dramatically complicate the form of the result, as compared to \eqref{f34}. Taking the
well-known dispersion functions for the eigenmodes of the Lifshitz problem, one can easily exemplify those cumbersome 
expressions with frequency derivatives of the permittivities under the sign of integration along the real frequency 
half-axis. All these demonstrate that the representation of the free energy in the form \eqref{f34} is far more 
preferable as compared to \eqref{f40}, \eqref{f41}.

Regarding the quantity $\rho(\omega)$, one notes that it enters \eqref{f37} as an effective spectral density of the
excitations. With $\rho(\omega)$ reducing to a delta-function in the limit of negligibly small dissipation, one easily 
gets from \eqref{f37} the free energy of a free oscillator. At the same time, the quantity $\Delta_\lambda\rho(\omega)$ 
in \eqref{f36} and \eqref{f41} is, at best, the difference between two effective spectral densities of the system taken
for the states with and without the interaction considered. 

Furthermore, the expressions for different thermodynamic potentials will only contain the same quantity 
$\Delta_\lambda\rho(\omega)$ in the absence of its temperature dependence. According to the general theory of 
Casimir and van der Waals forces, formulas \eqref{f34} and \eqref{f40}, \eqref{f41} for the free energy remain valid in 
the presence of the temperature dependent permittivities, that leads to a temperature dependence of 
$\Delta_\lambda\rho(\omega,T)$. In this case the expression for the internal energy, analogous to the free energy 
\eqref{f40}, \eqref{f41} will contain a different quantity 
$\Delta_\lambda\rho_E(\omega,T)\ne\Delta_\lambda\rho(\omega,T)$:
\bal
&\Delta_\lambda E=\int_{0}^{\infty}\dfrac{\hbar\omega}{2}\coth\dfrac{\hbar\omega}{2T}\Delta_\lambda\rho_E(\omega,T)
d\omega, \label{f42}\\
&\Delta_\lambda\rho_E(\omega,T)= \nonumber \\
&-\dfrac1{\pi}\Img \left(\dfrac{\p}{\p\omega}+\dfrac{T}{\omega}\dfrac{\p}{\p T}\right)
\int\rho(\beta)d\beta\ln D(\beta,\omega,T). \label{f43}
\eal

For the bilinear coupling of the system oscillator with the heat bath, the oscillator's damping function in the
Zwanzig-Caldeira-Leggett model is known to be independent of the temperature. The extension of the model, which results 
in the same formulas \eqref{f25}, \eqref{f26}, \eqref{f33} and \eqref{f35}-\eqref{f39}, but in the presence of a
temperature dependent damping function $\gamma(\omega,T)$ induced by the bilinear coupling with the environment, will be
developed in Section \ref{sec: freeenmod}.

\section{Pair and many-body interactions between different parts of the heat bath}
\label{sec: gen3}

After identifying a close analogy between the two theories, one can specify the counterpart in the
model of a damped oscillator that would correspond, at least formally, to the Casimir and van der Waals interaction
between macroscopic bodies. As will be demonstrated in the present section, the interaction between different parts of 
the heat bath is the counterpart in question. 

Although the environmental constituents do not affect each other directly, there occurs an effective indirect 
interaction between them through their coupling with the system oscillator. As a consequence of the many-body origin of
the interaction, the free energy \eqref{f25} and other thermodynamic potentials are nonadditive with respect to 
individual constituents' contributions, whereas the damping function \eqref{f19} and the interaction spectral density 
\eqref{f20} are the additive quantities. 

The pair and the three-body components of the free energy follow from \eqref{f25} in the weak coupling limit. Let the 
heat bath be subdivided into a number of distinct parts, and the coresponding damping function is
$\gamma=\sum_{n=1}^{p}\gamma_n$. If the weak coupling limit applies to the interaction of the environment with the 
system oscillator, one can expand \eqref{f25} in powers of $\gamma$ and represent the 
free energy in the form $F=\sum_n F_n +\sum_{n<m}F_{nm}+\sum_{n<m<l}F_{n,m,l}+\dots$\,. Here the quantity $F_n$ is an 
individual contribution of the $n$th part of the environment to the free energy, containing only the parameter 
$\gamma_n$ and its powers. 

The pair interaction induces the pair term in the free energy, and the main contribution from
each pair corresponds to the ``attraction'' between the parts:
\be
F_{12}= -\, T\sum_{n=1}^{\infty}
\dfrac{\omega_n^2\gamma_1(i\omega_n)\gamma_2(i\omega_n)}{(\Omega^2+\omega_n^2)^2}.
\label{f44}
\ee
Similarly, the main three-body contribution to the free energy from the combination of three different parts is
\be
F_{123}= 2T\sum_{n=1}^{\infty}
\dfrac{\omega_n^3\gamma_1(i\omega_n)\gamma_2(i\omega_n)\gamma_3(i\omega_n)}{(\Omega^2+\omega_n^2)^3}.
\label{f45}
\ee

The sums over the Matsubara frequencies in \eqref{f44}, \eqref{f45} converge even in the simplest Ohmic regime, when 
they can be expressed via the elementary and special functions, respectively. Thus one finds for the pair contribution,
assuming $\gamma_{1,2}$ to be independent of frequency,
\be
F_{12}=-\dfrac{2\hbar\gamma_1\gamma_2}{\Omega}\left[\coth\dfrac{\hbar\Omega}{2T}-
\dfrac{\hbar\Omega}{2T}\sinh^{-2}\dfrac{\hbar\Omega}{2T}\right].
\label{f46}
\ee

Vanishing interaction in the high temperature limit, that in general does not take place, is associated here with the 
same specific property of the Zwanzig-Caldeira-Leggett model that was mentioned in Section \ref{sec: freeen}.

\section{Damped oscillator linearly interacting with the dissipative environment}
\label{sec: freeenmod}

A large number of free oscillators with infinitesimal damping functions, that describe the spectral structure of the 
heat bath in the damped oscillator model, can be considered as representing environmental eigenmodes. The corresponding 
eigenfrequencies are real, since the heat bath in the absence of the system oscillator is assumed to be a closed system
itself.

The coupling of the system oscillator with an individual mode is assumed to be weak, which explains the very small 
perturbation of the environmental constituents, and generally supports the bilinear form of the interaction Hamiltonian. At the 
same time, the system oscillator can experience a pronounced dissipative influence collectively produced by a large
number of environmental modes. As a result, in the original model the frequency dependent finite damping function 
$\gamma(\omega)$ of the system oscillator can vary substantially, in accordance with \eqref{f19}-\eqref{f21}, when the 
interaction spectral density changes together with the spectral distribution of environmental real eigenfrequencies
\cite{CaldeiraLeggett1983,GrabertSchrammIngold1988,Weiss2012,SprengIngoldWeiss2013,AdamietzIngoldWeiss2014}.

Such an approach is known to be widely and efficiently used for studying problems that mostly focus on the emergence of
dissipation and its effects in quantum dynamical systems. On the other hand, one cannot exclude a possibility that the 
modes coupled to the system oscillator do not present all environmental degrees of freedom. Additional internal channels
of the heat bath, not linked directly to the system oscillator but, nevertheless, in contact with the basic 
environmental modes, can result in frequency and temperature dependent dissipative susceptibilities
$\chi_\alpha(\omega,T)$ that replace those with infinitesimal damping functions of the original model. A system 
oscillator's influence on the environmental susceptibilities, assumed to be negligible, should be compared in this case 
with the corresponding contributions of the additional channels that could dominate it.

With an unspecified part of the total Hamiltonian associated with the additional environmental channels, the extended
model outlined here is, generally speaking, not intended to study the total system's quantum dynamics. However, as shown
in this section, the interaction-induced thermodynamic quantities of the damped oscillator can be described within the 
extended framework analogously to the case considered in Section \ref{sec: freeen}. The key point here is that the 
interaction part of the extended Hamiltonian, which is required for obtaining the thermodynamic quantities, remains 
unchanged and can be separated out from the total Hamiltonian by taking its derivative over the corresponding coupling 
parameter.

Thus, one represents the total Hamiltonian in the form
\bml
\hat{H}=\dfrac{\hat{P}^2}{2M}+\dfrac12 M\Omega^2 \hat{Q}^2 -\hat{Q}f_{\text{ext},Q}-
\sum_{\alpha=1}^N\hat{q}_\alpha f_{\text{ext},\alpha}\\  + \sum_{\alpha=1}^N
\dfrac1{2\chi_\alpha(0,T)}\left(\hat{q}_\alpha-C_\alpha\chi_\alpha(0,T)\hat{Q}\right)^2+\hat{\widetilde{H}},
\label{f47}
\eml
where the term $\hat{\widetilde{H}}$ is not specified here, except it contains neither the system oscillator's 
operators $\hat{Q}$, $\hat{P}$, nor the coupling constants $C_\alpha$. 

The dynamic equation for $\hat{q}_\alpha(t)$, as follows from the total Hamiltonian \eqref{f47} after performing the 
Fourier transform, statistical averaging and the exclusion of additional degrees of freedom, is supposed to take the 
form
\be
\chi_{\alpha}^{-1}(\omega,T)q_\alpha(\omega)=C_\alpha Q(\omega)+f_{\text{ext},\alpha}(\omega),
\label{f48}
\ee
in accordance with the bilinear form of the interaction term and the definition of linear susceptibilities.
Eq. \eqref{f48} differs from \eqref{f14} solely due to the difference between \eqref{f11} and the dissipative 
environmental susceptibility $\chi_{\alpha}(\omega,T)$ of the $\alpha$th constituent of the heat bath.

Following Section \ref{sec: model} with Eqs. \eqref{f47}, \eqref{f48} instead of \eqref{f12}, \eqref{f14}, one arrives 
at the same equalities \eqref{f1}-\eqref{f3} and fluctuation-dissipation relationships \eqref{f4}-\eqref{f7}, as well as
\eqref{f9}, \eqref{f10} for the system damped oscillator. One can also conclude that the equalities \eqref{f13}, 
\eqref{f15}, \eqref{f16} and \eqref{f19}-\eqref{f21} will be satisfied with the modified susceptibilities 
$\chi_{\alpha}(\omega,T)$ substituted for the original ones \eqref{f11}. 

Going on to the derivation in Section \ref{sec: freeen}, one multiplies the coupling constants $C_\alpha$ in 
the Hamiltonian \eqref{f47} by the same coupling parameter $\lambda$. This does not affect the environmental
susceptibilities $\chi_{\alpha}(\omega,T)$, that are not noticeably influenced by the system oscillator and, therefore, 
not dependent on $\lambda$. On account of \eqref{f19}, the system oscillator's damping function is the quadratic 
function on $\lambda$ as before: $\gamma(\omega,T,\lambda)=\lambda^2\gamma(\omega,T)$. Thus, after obtaining the same 
expression \eqref{f23} for the derivative of the modified Hamiltonian \eqref{f47} over $\lambda$ and using the 
relationships \eqref{f4}, \eqref{f5} and \eqref{f15}, \eqref{f16}, one can integrate over the coupling parameter and 
obtain the same expressions \eqref{f24}-\eqref{f26}, \eqref{f33} for the interaction-induced free energy of the damped 
system oscillator, as well as formula \eqref{f27} for the reduced partition function. 

The free energy written in any of the forms \eqref{f25}, \eqref{f26}, \eqref{f33}, \eqref{f37}-\eqref{f39} now applies
to the case, in which the oscillator's damping function $\gamma(\omega,T)$ depends, in any admissible form, not only on
the frequency but also on the temperature, despite the bilinear interaction of the system oscillator with the heat 
bath. However, the expression \eqref{f28} for the internal energy of the damped oscillator is to be modified under such
conditions:
\be
E\!=\!T\sideset{}{'}\sum_{n=0}^{\infty}\dfrac{2\Omega^2\!+\omega_n\gamma(i\omega_n)\!-\!
\omega_n^2\!\left(\!\frac{\p}{\p\omega_n}\!+\!\frac{T}{\omega_n}\frac{\p}{\p T}\!\right)\!\gamma(i\omega_n,T)}{\Omega^2+
\omega_n^2+\omega_n\gamma(i\omega_n,T)}.
\label{f49}
\ee 

Eq.\eqref{f49} can also be represented in the form 
\bal
&E=\int_{0}^{\infty}\dfrac{\hbar\omega}{2}\coth\dfrac{\hbar\omega}{2T}\rho_E(\omega,T)d\omega, \label{f50}\\
&\rho_E(\omega,T)= -\dfrac1{\pi}\Img \left(\dfrac{\p}{\p\omega}+\dfrac{T}{\omega}\dfrac{\p}{\p T}\right)
\ln \bigl(MD(\omega,T)\bigr), \label{f51}
\eal
that agrees with \eqref{f42}, \eqref{f43}.

The extended model presented in this section demonstrates that thermodynamics of the quantum damped oscillator, 
bilinearly interacting with the heat bath, can be theoretically described without the model-imposed assumptions 
simplifying the realistic structure of the environment. This, in particular, makes it possible for the RCL circuit with 
metal resistance, which is controlled by the electron-phonon interaction, the electron scattering by impurities, and 
other internal physical processes in the metal, to be studied within the extended model. The inclusion of model-free 
internal environmental processes into the model of a damped oscillator has been introduced here along the lines of the 
general theory of Casimir and van der Waals interactions, in its application to realistic condensed matter systems 
with permittivities of a general form. 

The above consideration also demonstrates the intrinsic inconsistency of recent criticism of the general theory of
Casimir and van der Waals forces, which is based on the absence in the theory of an explicit expression for the total 
Hamiltonian of the quantum electromagnetic field in the condensed matter systems \cite{RosaDalvitMilonni2010,%
Barton2010,Philbin2010,RosaDalvitMilonni2011,Philbin2011}. The conclusion made in Refs. \cite{RosaDalvitMilonni2010,%
Barton2010,Philbin2010,RosaDalvitMilonni2011,Philbin2011} is that such a theory is incapable of presenting a consistent
microscopic quantum description of thermodynamic quantities for interacting quantum systems. 

The extended model is a comparatively simple example that reveals the erroneous nature of this claim. It shows 
that there is no need to know the environmental part of the total Hamiltonian, which contains no corresponding coupling
constants, in order to obtain the interaction-induced thermodynamic quantities of the quantum system bilinearly 
interacting with the environment. As the fluctuation-dissipation relationships identify response functions based on 
their microscopic expressions, the approach has the firm microscopic basis. The environmental linear response functions 
can further be considered as known from other theoretical or experimental studies. This allows one to separate the given
problem from a full dynamic description of the total quantum system.

One should also mention a number of models for absorbing condensed media that interact with the quantum electromagnetic 
field, which, in particular, are used to study the Casimir and van der Waals interaction \cite{Kupiszewska1992,%
Barton1997,BuhmannWelsch2007,ScheelBuhmann2008,RosaDalvitMilonni2010,RosaDalvitMilonni2010a,Barton2010,Barton2011,%
RosaDalvitMilonni2011,Philbin2011,EberleinZietal2012a,Bennett2014,Bordag2017,Braun2017,KlattFariasDalvitBuhmann2017,%
Bordag2018,SafariBarcellonaBuhmannSalam2020}. Those models include linearly polarizable oscillators, or the medium 
consisting of microscopic harmonic fields, which interact with the electromagnetic field in the way analogous to 
Refs.~\cite{Fano1956,Hopfield1958}, and, on top of that, they interact with the oscillators of the heat bath 
\cite{Kupiszewska1992,HuttnerBarnett1992}. The latter interaction results in the dissipative effects in the system and 
has a certain similarity to the damped oscillator model considered in this paper. The canonical quantization of the 
electromagnetic field, carried out by Huttner and Barnett in Ref.~\cite{HuttnerBarnett1992} after a Fano diagonalization
\cite{Fano1961} of the total model Hamiltonian, has led to the quantum current noise that is compatible with all 
fluctuation-dissipation relations. In addition, the functional parameters of the interaction Hamiltonian were chosen to 
imitate any given frequency dependent dissipative medium dielectric function that satisfies the Kramers-Kronig 
relationships. The resulting macroscopic quantum electrodynamics in the absorbing dispersive media has allowed the 
quantum effects of the matter-field interaction in material bodies and their vicinity to be studied within the model 
\cite{BarnettHuttnerLoudon1992,HuttnerBarnett1992,GrunerWelsch1995,MatloobLoudonBarnettJeffers1995,MatloobLoudon1996,%
GrunerWelsch1996,DungKnoellWelsch1998,TipKnoellScheelWelsch2001,SuttorpWubs2004,BhatSipe2006,ScheelBuhmann2008,%
Philbin2010,EberleinZietal2012,BennettEberlein2013,Buhmann2013,Drezet2017,Kosiketal2020,Forestiereetal2020,%
YouNellikkaetal2020}. 

Considering the model results are ultimately expressed via the system's dielectric function with the arbitrary frequency
dependence, one might expect them to be of a more general character than the underlying model Hamiltonian. This would 
mean, however, that the final results can always be expressed via the permittivities of absorbing dispersive media. 
Although this point is valid in respect of the Casimir and van der Waals interaction 
\cite{DzyaloshinskiiPitaevskii1959}, it has not been confirmed in general \cite{PerelPinskii1968} (see also 
\cite{Pitaevskii2011}). Furthermore, unlike the frequency dependence, the permittivity's temperature dependence cannot 
be reproduced with the microscopic model Hamiltonian, which contains only quadratic and bilinear operator terms, with no
temperature dependent parameters. 

Thus, the results for dispersion interactions, obtained with the models for absorbing condensed media, can be expressed 
via material permittivities and must be in agreement with the general theory of Casimir and van der Waals forces 
\cite{DzyaloshinskiiPitaevskii1959,DzyaloshinskiiLifshitzPitaevskii1961} within its range of applicability. This, 
however, bears no direct relation to the results of the present paper. The interaction of medium oscillators with 
oscillators of the medium heat bath, that contributes to permittivities, is not dependent on the distance between the
condensed matter bodies. Therefore, a substantial part of free energy, induced by such an interaction, has no relevance 
to the distance dependent Casimir and van der Waals contribution to free energy. 

By contrast, the contribution to free 
energy examined in the paper is entirely induced by the interaction of the system oscillator with the environment. The
paper has uncovered the similarity between the damped oscillator model in its original form, where the system oscillator
interacts bilinearly with oscillators of the heat bath, and the general theory of Casimir and van der Waals forces, 
which is free from model assumptions when it considers the long wavelength fluctuational quantum electromagnetic field 
interacting with condensed matter bodies.

\section{Conclusions}
\label{sec: conclusions}

One can conclude that equilibrium thermodynamics of a quantum damped oscillator bilinearly coupled to the heat bath, 
studied within the Zwanzig-Caldeira-Leggett model, can be described directly by the general theory of Casimir and
van der Waals forces. A significant overlap between the two theories is identified and specified above. A model for 
a damped oscillator, which extends the original model and allows frequency and temperature dependence of the system 
oscillator's damping function, is presented together with the modified corresponding thermodynamic potentials.


\begin{thebibliography}{99}

\bibitem{Lifshitz1955}
E.~M.~Lifshitz, Sov. Phys. JETP {\bf 2}, 73 (1956).

\bibitem{DzyaloshinskiiPitaevskii1959}
I.~E.~Dzyaloshinskii and  L.~P.~Pitaevskii, Sov. Phys. JETP {\bf 9}, 1282 (1959).

\bibitem{DzyaloshinskiiLifshitzPitaevskii1959}
I.~E.~Dzyaloshinskii, E.~M.~Lifshitz, and  L.~P.~Pitaevskii, Sov. Phys. JETP {\bf 10}, 161 (1959).

\bibitem{DzyaloshinskiiLifshitzPitaevskii1961}
I.~E.~Dzyaloshinskii, E.~M.~Lifshitz, and  L.~P.~Pitaevskii, Adv. Phys. {\bf 10}, 165 (1961). 

\bibitem{Lamoreaux2004} S.~K.~Lamoreaux, Rep. Prog. Phys. {\bf 68}, 201 (2004).

\bibitem{Parsegian2006} V.~A.~Parsegian, {\it Van der Waals Forces: A Handbook for Biologists, Chemists, Engineers, and 
Physicists} (Cambridge Univ. Press, Cambridge, UK, 2006).

\bibitem{KlimchitskayaMohideenMostepanenko2009} G.~L.~Klimchitskaya, U.~Mohideen, and V.~M.~Mostepanenko, 
Rev. Mod. Phys. {\bf 81}, 1827 (2009).

\bibitem{Bordagetal2009} M.~Bordag, G.~L.~Klimchitskaya, U.~Mohideen, and 
V.~M.~Mostepanenko, {\it  Advances in the Casimir Effect} (Oxford Univ. Press, Oxford, UK, 2009).

\bibitem{FrenchParsegianPodgorniketal2010} R.~H.~French, V.~A.~Parsegian, R.~Podgornik, R.~F.~Rajter, A.~Jagota, 
J.~Luo, D.~Asthagiri, M.~K.~Chaudhury, Y.-m.~Chiang, S.~Granick, S.~Kalinin, M.~Kardar, R.~Kjellander, D.~C.~Langreth, 
J.~Lewis, S.~Lustig, D.~Wesolowski, J.~S.~Wettlaufer, W.-Y.~Ching, M.~Finnis, F.~Houlihan, O.~A.~von~Lilienfeld, 
C.~J.~van~Oss, and T.~Zemb, Rev.~Mod.~Phys. {\bf 82}, 1887 (2010).

\bibitem{Dalvitetal2011} {\it Casimir Physics}, Vol. 834 of Lecture Notes Phys., ed. by D.~Dalvit, P.~Milonni, 
D.~Roberts, and F.~Rosa (Springer-Verlag, Berlin/Heidelberg, 2011).

\bibitem{Woodsetal2016} L.~M.~Woods, D.~A.~R.~Dalvit, A.~Tkatchenko,
P.~Rodriguez-Lopez, A.~W.~Rodriguez, and R.~Podgornik, Rev. Mod. Phys. {\bf 88}, 045003 (2016).

\bibitem{FordLewisOConnell1985} 
G.~W.~Ford, J.~T.~Lewis, and R.~F.~O'Connell, Phys. Rev. Lett. {\bf 55}, 2273 (1985).

\bibitem{FordLewisOConnell1988} G.~Ford, J.~Lewis, and R.~O'Connell, Ann. Physics {\bf 185}, 270 (1988).

\bibitem{FordLewisOConnell1988_2} G.~W.~Ford, J.~T.~Lewis, and R.~F.~O'Connell, J. Stat. Phys. {\bf 53}, 439 (1988).

\bibitem{FordOConnell2005} G.~W.~Ford and R.~F.~O'Connell, Physica E {\bf 29}, 82 (2005).

\bibitem{HaenggiIngold2006} P.~H\"anggi and G.-L.~Ingold, Acta Phys. Polon. B {\bf 37}, 1537 (2006).

\bibitem{FordOConnell2007} G.~W.~Ford and R.~F.~O'Connell, Phys. Rev. B {\bf 75}, 134301 (2007).

\bibitem{HoerhammerBuettner2008} C.~H\"orhammer and H.~B\"uttner, J. Stat. Phys. {\bf 133}, 1161 (2008).

\bibitem{HaenggiIngoldTalkner2008} P.~H\"anggi, G.-L.~Ingold, and P.~Talkner, New J. Phys. {\bf 10}, 115008 (2008).

\bibitem{IngoldHaenggiTalkner2009} G.-L. Ingold, P.~H\"anggi, and P.~Talkner, Phys. Rev. E {\bf 79}, 061105 (2009).

\bibitem{Philbin2012} T.~G.~Philbin, New J. Phys. {\bf 14}, 083043 (2012).

\bibitem{SprengIngoldWeiss2013} B.~Spreng, G.-L.~Ingold, and U.~Weiss, EPL {\bf 103}, 60007 (2013).

\bibitem{AdamietzIngoldWeiss2014} R.~Adamietz, G.-L.~Ingold, and U.~Weiss, Eur. Phys. J. B {\bf 87}, 90 (2014).

\bibitem{PhilbinAnders2016} T.~G.~Philbin and J.~Anders, J. Phys. A {\bf 49}, 215303 (2016).

\bibitem{KolarRyabovFilip2019} M.~Kol\'a{\v{r}}, A.~Ryabov, and R.~Filip, Sci. Rep. {\bf 9}, 10855 (2019).

\bibitem{TalknerHaenggi2020} P.~Talkner and P.~H\"anggi, Rev. Mod. Phys. {\bf 92}, 041002 (2020).

\bibitem{Magalinskii1959} V.~B.~Magalinskii, Sov. Phys. JETP {\bf 9}, 1381 (1959).

\bibitem{FordKacMazur1965} G.~W.~Ford, M.~Kac, and P.~Mazur, J. Math. Phys, {\bf 6}, 504 (1965).

\bibitem{Ullersma1966_1_4} P.~Ullersma, Physica {\bf 32}, 27, 56, 74, 90 (1966).

\bibitem{Zwanzig1973} R.~Zwanzig, J. Stat. Phys, {\bf 9}, 215 (1973).

\bibitem{CaldeiraLeggett1981} A.~O.~Caldeira and A.~J.~Leggett, Phys. Rev. Lett. {\bf 46}, 211 (1981).

\bibitem{Schmid1982} A.~Schmid, J. Low Temp. Phys. {\bf 49}, 609 (1982).

\bibitem{CaldeiraLeggett1983} A.~O.~Caldeira and A.~J.~Leggett, Annals of Physics {\bf 149}, 374 (1983).

\bibitem{GrabertWeissTalkner1984} H.~Grabert, U.~Weiss, and P.~Talkner, Z. Phys. B {\bf 55}, 87 (1984).

\bibitem{RiseboroughHaenggiWeiss1985} P.~S.~Riseborough, P.~Hanggi, and U.~Weiss, Phys. Rev. A {\bf 31}, 471 (1985).

\bibitem{FordKac1987} G.~W.~Ford and M.~Kac, J. Stat. Phys. {\bf 46}, 803 (1987).

\bibitem{GrabertSchrammIngold1988} H.~Grabert, P.~Schramm, and G.-L.~Ingold, Phys. Rep. {\bf 168}, 115 (1988).

\bibitem{FordLewisOConnell1988_3} G.~W.~Ford, J.~T.~Lewis, and R.~F.~O'Connell, Phys. Rev. A {\bf 37}, 4419 (1988).

\bibitem{HaenggiIngold2005} P.~H\"anggi and G.-L.~Ingold, Chaos {\bf 15}, 026105 (2005).

\bibitem{Weiss2012} U.~Weiss, {\it Quantum Dissipative Systems}, 4th ed., World Scientific, Singapore (2012).

\bibitem{Caldeira2014} A.~O.~Caldeira, {\it An Introduction to Macroscopic Quantum Phenomena and Quantum Dissipation}
(Cambridge Univ. Press, New York, 2014).

\bibitem{Leggettetal1987} A.~J.~Leggett, S.~Chakravarty, A.~T.~Dorsey, M.~P.~A.~Fisher, A.~Garg, and W.~Zwerger, 
Rev. Mod. Phys. {\bf 59}, 1 (1987).

\bibitem{IngoldLambrechtReynaud2009} G.-L.~Ingold, A.~Lambrecht, and S.~Reynaud, Phys. Rev. E {\bf 80}, 041113 (2009).

\bibitem{BarashGinzburg1972} Yu.~S.~Barash and V.~L.~Ginzburg, JETP Lett. {\bf 15}, 403 (1972).

\bibitem{footnote} Following the analogy identified here between the results of the two theories, the dissipative 
material properties should be compared solely with the dissipative properties of the heat bath rather than the system 
oscillator. The role of the latter in the model is analogous to that played by the fluctuational electromagnetic field 
in the theory of Casimir and van der Waals forces.

\bibitem{Obcemea1987} C.~H.~Obcemea, Int. J. Quant. Chem. {\bf 31}, 113 (1987).

\bibitem{FLondon1937} F.~London, Trans. Faraday Soc. {\bf 33}, 8b (1937).

\bibitem{Casimir1948} H.~B.~G.~Casimir, Proc. Kon. Nederl. Akad. Wetensch. {\bf 51}, 793 (1948).

\bibitem{vanKampenNijboerSchram1968} N.~Van Kampen, B.~Nijboer, and K.~Schram, Phys. Lett. A {\bf 26}, 307 (1968).

\bibitem{NinhamParsegianWeiss1970} B.~W.~Ninham, V.~A.~Parsegian, and G.~H.~Weiss, J. Stat. Phys. {\bf 2}, 323 (1970).

\bibitem{BarashGinzburg1975} Y.~S.~Barash and V.~L.~Ginzburg, Sov. Phys. Usp. {\bf 18}, 305 (1975).

\bibitem{AbrikosovGorkovDzyaloshinskii1965} A.~A.~Abrikosov, L.~P.~Gor'kov, I.~Y.~Dzyaloshinskii, 
{\it Quantum Field Theoretical Methods in Statistical Physics}, 2nd ed. (Pergamon, Oxford, 1965).

\bibitem{Lifshitz1995} E.~M.~Lifshitz, L.~P.~Pitaevskii, {\it Statistical Physics.} Part 2. 
{\it Theory of the Condensed State} (Butterworth-Heinemann, Oxford, 1995).

\bibitem{Barash1988} Y. S. Barash, {\it Van der Waals Forces} (Nauka, Moscow, 1988) [in Russian].

\bibitem{BarashGinzburg1989} Y.~S.~Barash and V.~L.~Ginzburg, in {\it The Dielectric Function of Condensed Systems},
Vol. 24 of {Modern Problems in Condensed Matter Sciences}, Ed. by L.~V.~Keldysh, D.~A.~Kirzhnitz, and A.~A.~Maradudin, 
(Elsevier, Amsterdam, 1989), Chap. 6, p. 389.

\bibitem{LandauLifshitz5_1980} L.~Landau and E.~Lifshitz, {\it Course of Theoretical Physics}, Vol. 5: 
{\it Statistical Physics}, 3rd ed. (Butterworth-Heinemann, Oxford, 1980).

\bibitem{Abrikosov1988} A.~A.~Abrikosov, {\it Fundamentals of the Theory of Metals} (North-Holland, Amsterdam, 1988).

\bibitem{footnote2} That had been done by Lifshitz before the Matsubara's paper was published.

\bibitem{RosaDalvitMilonni2010} F.~S.~S.~Rosa, D.~A.~R.~Dalvit, and P.~W.~Milonni, Phys. Rev. A {\bf 81}, 033812 (2010).

\bibitem{Barton2010} G.~Barton, New J. Phys. {\bf 12}, 113045 (2010).

\bibitem{Philbin2010} T.~G.~Philbin, New J. Phys. {\bf 12}, 123008 (2010).

\bibitem{RosaDalvitMilonni2011} F.~S.~S.~Rosa, D.~A.~R. ~Dalvit, and P.~W.~Milonni, 
Phys. Rev. A {\bf 84}, 053813 (2011).

\bibitem{Philbin2011} T.~G.~Philbin, New J. Phys. {\bf 13}, 063026 (2011).

\bibitem{Kupiszewska1992} D.~Kupiszewska, Phys. Rev. A 46, 2286 (1992).

\bibitem{Barton1997} G.~Barton, Proc. R. Soc. Lond. A {\bf 453}, 2461 (1997).

\bibitem{BuhmannWelsch2007} S.~Y. Buhmann and D.-G.~Welsch, Prog. Quantum Electron. {\bf 31}, 51 (2007).

\bibitem{ScheelBuhmann2008} S.~Scheel and S.~Y.~Buhmann, Acta Phys. Slov. {\bf 58}, 675 (2008).

\bibitem{RosaDalvitMilonni2010a} F.~S.~S.~Rosa, D.~A.~R.~Dalvit, and P.~W.~Milonni, in {\it Doing Physics: 
A Festshcrift for Thomas Erber}, ed. by P.~W.~Johnson (Illinois Inst. Technol. Press, Chicago, IL, 2010), Chap. 18, 
p. 187; arXiv:0912.0279.

\bibitem{Barton2011} G.~Barton, J. Phys.: Condens. Matter {\bf 23}, 355004 (2011).

\bibitem{EberleinZietal2012a} C.~Eberlein and R.~Zietal, Phys. Rev. A {\bf 86}, 062507 (2012).

\bibitem{Bennett2014} R.~Bennett, Phys. Rev. A {\bf 89}, 062512 (2014).

\bibitem{Bordag2017} M.~Bordag, Phys. Rev. A {\bf 96}, 062504 (2017).

\bibitem{Braun2017} M.~A.~Braun, Theor. Math. Phys. {\bf 190}, 237 (2017).

\bibitem{KlattFariasDalvitBuhmann2017} J.~Klatt, M.~B.~Far\'ias, D.~A.~R. Dalvit, and S.~Y.~Buhmann, 
Phys. Rev. A {\bf 95}, 052510 (2017).

\bibitem{Bordag2018} M.~Bordag, Theor. Math. Phys. {\bf 195}, 834 (2018).

\bibitem{SafariBarcellonaBuhmannSalam2020} H.~Safari, P.~Barcellona, S.~Y.~Buhmann, and A.~Salam, 
New J. Phys. {\bf 22}, 053049 (2020).

\bibitem{Fano1956} U.~Fano, Phys. Rev. {\bf 103}, 1202 (1956).

\bibitem{Hopfield1958} J.~J.~Hopfield, Phys. Rev. {\bf 112}, 1555 (1958).

\bibitem{HuttnerBarnett1992} B.~Huttner and S.~M.~Barnett, Phys. Rev. A {\bf 46}, 4306 (1992).

\bibitem{Fano1961} U.~Fano, Phys. Rev. {\bf 124}, 1866 (1961).

\bibitem{BarnettHuttnerLoudon1992} S.~M.~Barnett, B.~Huttner, and R.~Loudon, Phys. Rev. Lett. {\bf 68}, 3698 (1992).

\bibitem{GrunerWelsch1995} T.~Gruner and D.-G.~Welsch, Phys. Rev. A {\bf 51}, 3246 (1995).

\bibitem{MatloobLoudonBarnettJeffers1995} R.~Matloob, R.~Loudon, S.~M.~Barnett, and J.~Jeffers, 
Phys. Rev. A {\bf 52}, 4823 (1995).

\bibitem{MatloobLoudon1996} R.~Matloob and R.~Loudon, Phys. Rev. A {\bf 53}, 4567 (1996).

\bibitem{GrunerWelsch1996} T.~Gruner and D.-G.~Welsch, Phys. Rev. A {\bf 53}, 1818 (1996).

\bibitem{DungKnoellWelsch1998} H.~T.~Dung, L.~Kn\"oll, and D.-G.~Welsch, Phys. Rev. A {\bf 57}, 3931 (1998).

\bibitem{TipKnoellScheelWelsch2001} A.~Tip, L.~Kn\"oll, S.~Scheel, and D.-G.~Welsch, 
Phys. Rev. A {\bf 63}, 043806 (2001).

\bibitem{SuttorpWubs2004} L.~G.~Suttorp and M.~Wubs, Phys. Rev. A {\bf 70}, 013816 (2004).

\bibitem{BhatSipe2006} N.~A.~R.~Bhat and J.~E.~Sipe, Phys. Rev. A {\bf 73}, 063808 (2006).

\bibitem{EberleinZietal2012} C.~Eberlein and R.~Zietal, Phys. Rev. A {\bf 86}, 022111 (2012).

\bibitem{BennettEberlein2013} R.~Bennett and C.~Eberlein, Phys. Rev. A {\bf 88}, 012107 (2013).

\bibitem{Buhmann2013} S.~Y.~Buhmann, {\it Dispersion Forces I: Macroscopic Quantum Electrodynamics and Ground-State 
Casimir, Casimir–Polder and van der Waals Forces}, Vol. 247 of Springer Tracts Mod. Phys. (Springer, Berlin, 
Heidelberg, 2013).

\bibitem{Drezet2017} A.~Drezet, Phys. Rev. A {\bf 96}, 033849 (2017).

\bibitem{Kosiketal2020} M.~Kosik, O.~Burlayenko, C.~Rockstuhl, I.~Fernandez-Corbaton, and K.~S{\l}owik, 
Sci. Rep. {\bf 10}, 10855 (2020).

\bibitem{Forestiereetal2020} C.~Forestiere, G.~Miano, M.~Pascale, and R.~Tricarico, 
Phys. Rev. A {\bf 102}, 043704 (2020).

\bibitem{YouNellikkaetal2020} C.~You, A.~C.~Nellikka, I.~D.~Leon, and O.~S.~Maga{\~{n}}a-Loaiza,  
Nanophotonics {\bf 9}, 1243 (01 Jun. 2020).

\bibitem{PerelPinskii1968} V.~I.~Perel' and Y.~M.~Pinskii, Sov. Phys. JETP {\bf 27}, 1014 (1968).

\bibitem{Pitaevskii2011} L.~P.~Pitaevskii, in Ref. [10], Chap. 2,  Lect. Notes Phys. {\bf 834}, 23 (2011).


\end{thebibliography}
\end{document}